# Distinct Electronic Structure for the Extreme Magnetoresistance in YSb

Junfeng He[1, 2, #], Chaofan Zhang[1, 2, #], Nirmal J. Ghimire[3], Tian Liang[1, 2], Chunjing Jia[1, 2], Juan Jiang[4, 5, 6], Shujie Tang[1, 2], Sudi Chen[1, 2], Yu He[1, 2], S.-K. Mo[4], C. C. Hwang[6], M. Hashimoto[7], D. H. Lu[7], B. Moritz[1, 2], T. P. Devereaux[1, 2], Y. L. Chen[5, 8], J. F. Mitchell[3] & Z.-X. Shen[1, 2*]

[1]*Stanford Institute for Materials and Energy Sciences, SLAC National Accelerator Laboratory, 2575 Sand Hill Road, Menlo Park, California 94025, USA*

[2]*Geballe Laboratory for Advanced Materials, Departments of Physics and Applied Physics, Stanford University, Stanford, California 94305, USA*

[3]*Materials Science Division, Argonne National Laboratory, Argonne, Illinois 60439, USA*

[4]*Advanced Light Source, Lawrence Berkeley National Laboratory, Berkeley, California 94720, USA*

[5]*School of Physical Science and Technology, ShanghaiTech University, Shanghai 200031, P. R. China*

[6]*Pohang Accelerator Laboratory, Pohang University of Science and Technology, Pohang 790-784, Korea*

[7]*Stanford Synchrotron Radiation Lightsource, SLAC National Accelerator Laboratory, 2575 Sand Hill Road, Menlo Park, California 94025, USA*

[8]*Physics Department, Oxford University, Oxford, OX1 3PU, UK*

abstract
An extreme magnetoresistance (XMR) has recently been observed in several non-magnetic semimetals. Increasing experimental and theoretical evidence indicates that the XMR can be driven by either topological protection or electron-hole compensation. Here, by investigating the electronic structure of a XMR material, YSb, we present spectroscopic evidence for a special case which lacks topological protection and perfect electron-hole compensation. Further investigations reveal that a cooperative action of a substantial difference between electron and hole mobility and a moderate carrier compensation might contribute to the XMR in YSb.

The recent discovery of XMR in several non-magnetic semimetals has led to considerable efforts directed toward understanding its mechanism and exploring potential applications[1-14]. The existence of large and non-saturating magnetoresistance (MR) differentiates these materials from typical metals, where only small MR is expected. Among various candidate mechanisms that might contribute to XMR in these semimetals, two scenarios have arguably obtained the most attention[1,2,4-6,8-11,13]. One is a novel topological protection mechanism which suppresses backscattering in the compounds at zero magnetic field. The lifting of the topological protection by external magnetic field gives rise to the XMR[4-6]. This scenario relies entirely on the topologically non-trivial electronic structure in the XMR semimetals. For example, TaAs, NbP and NbAs are Weyl semimetals identified by Weyl nodes and Fermi arcs[15-19]; and $Cd_3As_2$ is a 3D Dirac semimetal characterized by the linear band dispersion[20]. The other scenario with increasing experimental and theoretical support[1,2,8,9,11,13] is based on a classical carrier compensation picture[21], in which the high MR is attributed to a balanced concentration of electrons and holes.

In this paper, we study the electronic structure of YSb, a new member of the XMR family, via a combination of first principle calculations and angle-resolved photoemission spectroscopy (ARPES) measurements. The simple rock-salt structure of YSb[22] provides an ideal platform for the theoretical investigation and the sharp band dispersion observed by ARPES enables a quantitative determination of the electronic structure. A general agreement has been seen between the calculations and measurements, clearly establishing that no topologically non-trivial electronic state is present in YSb. Therefore, the topological protection cannot account for the XMR in this material. Both electron and hole pockets are observed at the Fermi energy, but the electron/hole concentration ratio (~0.81 at 10K), estimated by the experimentally measured volume of electron and hole pockets, deviates from perfect carrier compensation. These observations make YSb a special non-magnetic semimetal which exhibits XMR but lacks topological protection and perfect electron-hole compensation. Further investigations reveal a third route: when a substantial difference between electron and hole mobility exists, its cooperative action with the moderate carrier compensation observed in YSb could contribute to XMR.

Single crystals of YSb were grown in Sb self-flux[22]. ARPES measurements were mainly carried out at Beamline 10.0.1 of the Advanced Light Source (ALS) of Lawrence Berkeley National Laboratory with a total energy resolution of ~15meV and a base pressure better than $5 \times 10^{-11}$ Torr. Preliminary experiments were performed at Beamline 5-4 of the Stanford Synchrotron Radiation Lightsource (SSRL) of SLAC National Accelerator Laboratory. *ab initio* electronic structure calculations were performed with the WIEN2k code package using the standard PBE-GGA exchange-correlation potential[23]. The modified Becke-Johnson (mBJ) potential[24] was incorporated for improved estimates of the band gaps.

YSb crystalizes in a face-centered cubic (FCC) structure whose 3D Brillouin zone is plotted in Fig. 1c. Its low energy electronic structure near the Fermi level is mainly characterized by two hole pockets at the zone center (Γ in Fig. 1c,f; see Fig. 2a and Fig. 3a,b for the detailed separation of the two pockets in momentum space) and an electron pocket at each zone corner (X in Fig. 1c,f and g). For simplicity, we hereafter label the hole pockets as α, β and the electron pocket as γ. The corresponding hole-like and electron-like bands are shown along two high symmetry directions in Fig. 1d,e (indicated by the dashed lines). We note that the number, momentum location, and shape of the Fermi pockets, as well as the overall band dispersions, all agree well with the theoretical calculations (compare Fig. 1d-g with a-c). Therefore, our calculations serve as a clear guide to understand the electronic structure of this material.

A direct observation from both first principles calculations and ARPES measurements is the absence of any non-trivial topological state in the electronic structure. Neither topological band inversion nor topological surface state is present in the calculated band structures (see Supplementary Fig. 1). Experimentally, no evidence of topologically non-trivial state (for example, Dirac linear dispersion, Dirac/Weyl nodes or Fermi arcs) is observed throughout the Brillouin zone (Fig. 1c-g). This makes YSb distinct from other XMR materials that crystallize in the rock-salt structure (for example, LaSb and LaBi), which have been proposed to be topological insulators[25]. On the other hand, the coexistence of both electron and hole pockets at Fermi energy seems to suggest that two types of carriers are contributing to the transport properties in YSb.

Before discussing the electron and hole pockets quantitatively, we note the appearance of several additional bands around Γ (two hole-like bands, marked as B1 and B2 in Fig. 1d; one electron-like band, marked as B3 in Fig. 2a). These bands show similar dispersion as the ones along W-X-W (Fig. 1b,e), but have much weaker spectral intensity. Therefore, an immediate interpretation is the projection of features along W-X-W due to the finite $K_z$ resolution in ARPES experiments. It has been reported that some features at different $K_z$ might coexist across the entire $K_z$ range when the 3D band dispersion of a material is sensitive to the $K_z$ resolution[26]. Alternatively, the additional bands may also come from band folding if a supermodulation of either electrons or lattice doubles the primitive cell of the crystal. This possibility is corroborated by the first principle calculations, which capture the additional bands by utilizing a doubled primitive cell (dashed lines in Fig. 1a). While no electronic order was reported in YSb, surface reconstruction is a typical supermodulation of the lattice on sample surface after cleaving[27]. Regardless of the particular origin, these bands do not contribute to the carrier counting in the Fermi pockets.

In order to estimate the electron and hole concentrations in YSb, detailed Fermi surface mappings have been carried out in both the $K_x$-$K_y$ plane at $K_z=4\pi c^{-1}$ (Fig. 2b, $K_z=4\pi c^{-1}$ is equal to $K_z=0$) and $K_x$-$K_z$ plane at $K_y=0$ (Fig. 2d). Continuous band evolution near Γ and X is shown in Fig. 2a (from cut1 to cut14 in Fig. 2b). Second derivative images with respect to the momentum are used to enhance the electronic structure. Both α and β hole-like bands cross the Fermi level around Γ (indicated by the black dashed lines in the panel for cut7), giving rise to two hole pockets around the zone center. As discussed above, the B3 electron-like band at Γ (being either a projection of features at another $K_z$ or a folded band) does not contribute to the estimation of carrier density via Luttinger volume[28]. The electronic structure near X is characterized by the γ electron-like band which forms an ellipsoidal electron pocket (Fig. 2a,b). This electron pocket is confirmed and better quantified by the measurements in the $K_x$-$K_z$ plane where the momentum cuts are parallel to a short axis of the ellipsoid (from cut1 to cut7 in Fig. 2d). Shown in Fig. 2c is the evolution of the γ band as a function of $K_z$, in which the dispersion of the electron-like band is well characterized (see also supplementary Fig. 2).

Fig. 3 shows the deduced Fermi surface area of the hole pockets in the $K_x$-$K_y$ plane at $K_z=0$ and that of the electron pocket in the $K_x$-$K_z$ plane at $K_y=0$. The electron pocket is well characterized by an ellipsoid of revolution (Fig. 1c,f and Fig. 3c) whose semi-principal axes length can be directly obtained from its projection on the $K_x$-$K_z$ plane at $K_y=0$ (Fig. 3c). The semi-major axis length is approximately 0.363 (Å$^{-1}$) and the two semi-minor axes are equal due to the crystal symmetry with an estimated length of ~ 0.0785 (Å$^{-1}$). We note that these numbers are consistent with those in the $K_x$-$K_y$ plane at $K_z=0$ (Fig. 1f and Fig. 2a), which confirms our determination of $K_z$. Considering three electron ellipsoids in one Brillouin zone (Fig. 1c, FCC lattice constant a=6.16 Å) yields an electron concentration of ~ 2.27 × 10$^{20}$(cm)$^{-3}$. The hole pockets are more complicated. While the β pocket is nearly spherical, the α pocket is stretched along each Γ-X direction in the 3D

Brillouin zone. A conservative estimation of the volume is to consider two spheres with different radii for each hole pocket (indicated by the blue and green dashed lines in Fig. 3a and b for α and β hole pockets respectively). The radius of the inner (outer) sphere is determined by the length of the semi-shortest (longest) axis. Therefore, the volume of the sphere sets a lower (upper) limit to the volume of each hole pocket. The estimated lower and upper limits to the hole concentration are ~ $2.72 \times 10^{20}(cm)^{-3}$ and ~ $4.31 \times 10^{20}(cm)^{-3}$ respectively. An alternative way to estimate the volume of the hole pockets is to compare with calculations. When the 2D projection of the calculated hole pockets matches that of the experiments in the $K_x$-$K_y$ plane at $K_z$=0, the calculation gives a good estimation of the 3D volume of the pockets (see supplementary Fig. 3). This method yields a hole concentration of ~ $2.81 \times 10^{20}(cm)^{-3}$ which lies in between the lower and upper limits. The resulting electron/hole concentration ratio is ~ 0.81, deviating from the perfect carrier compensation. We note this carrier concentration ratio does not change with temperature (see supplementary Fig. 4), distinct from that reported in $WTe_2$[2,9].

To investigate the origin of the XMR in YSb, we first consider the topological protection mechanism[4-6]. The prerequisite for this scenario is the existence of non-trivial topological states that can suppress the backscattering in the material. A direct way to identify the non-trivial topological states is to study their characteristic electronic features[15-20]. For example, topological insulators are characterized by the inversion of bulk bands and Dirac linear dispersion in the topological surface states; 3D Dirac semimetals are identified by the 3D linear dispersion in the bulk bands; Weyl semimetals are shown by the Weyl nodes and Fermi arcs. The absence of the characteristic electronic fingerprints for these non-trivial topological states in our ARPES measurements and first principles calculations indicates that the novel topological protection mechanism may not be essential to account for the XMR in YSb.

The second scenario to be considered is carrier compensation. The observation of both electron and hole pockets with very low carrier density (in the order of $10^{20}$ cm$^{-3}$) in YSb seems to be compatible with this picture. Nevertheless, a quantitative analysis yields an electron/hole concentration ratio of ~ 0.81, deviating from the perfect carrier compensation claimed in other XMR materials[1,2]. Therefore, a natural question is posed: whether the compensation scenario still contributes to XMR in YSb?

To address this question, we have performed a simulation using the standard two-band model[11]: MR= $\frac{n_e\mu_e n_h\mu_h(\mu_e+\mu_h)^2(\mu_0 H)^2}{(n_e\mu_e+n_h\mu_h)^2+(n_e-n_h)^2(\mu_e\mu_h)^2(\mu_0 H)^2}$, in which $n_e$, $n_h$ are the carrier concentration for electron and hole respectively; $\mu_e$, $\mu_h$ are the corresponding mobility for each carrier; and $\mu_0 H$ is the magnetic field. As shown in Fig. 4a, when electrons and holes are perfectly compensated ($n_e$=$n_h$), a quadratic MR is achieved with the absolute value similar to that obtained from transport measurements. For simplicity, the same electron and hole mobility is used in the simulation. As we expected, the MR is dramatically suppressed when the carrier concentration deviates from perfect compensation. The electron/hole concentration ratio in YSb ($n_e$/$n_h$~0.81) yields a saturated MR with the absolute value around two orders of magnitude smaller than that of the experiment (Fig. 4b vs. Fig. 4a). However, the suppression of MR due to the imperfect carrier compensation can be eased if there is a substantial difference between the electron and hole mobility. By fixing the $n_e$/$n_h$ ratio as 0.81 but using an electron mobility much larger than that of the hole [$\mu_e$=9 × 10$^5$ cm$^2$(Vs)$^{-1}$ and $\mu_h$=3.5 × 10$^3$ cm$^2$(Vs)$^{-1}$], we achieve an XMR which not only shows the correct absolute value but fits the curvature of the experimental result well (Fig. 4c).

Therefore, if a substantial mobility difference exists, its cooperative action with the observed moderate carrier compensation in YSb could give rise to the XMR. Indication of different carrier mobility in YSb is indeed observed by our transport measurements.

Following the typical method[4], the extrema in the Hall conductivity $\sigma_{xy}$ are used to determine the electron and hole mobility respectively ($\mu=1/|B|$), which yields $\mu_e \sim 6 \times 10^5$ cm$^2$(Vs)$^{-1}$ and $\mu_h \sim 4 \times 10^4$ cm$^2$(Vs)$^{-1}$ (see supplementary Fig.5). Nevertheless, we note the carrier mobility difference extracted from the Hall conductivity is not as large as the one needed in the simulation to fit the experimental XMR. Various factors might contribute to this discrepancy. First, the simple two band model itself may not be sufficient to fully describe a real material in a quantitative level. Second, a remarkable sample-to-sample variation in carrier mobility has been reported in XMR semimetals[4]. Third, it is possible that other driving mechanisms can coexist, such that the observed mobility difference does not account for the total contribution of MR in YSb. While more efforts are needed to quantitatively address these issues, a qualitative difference between the electron and hole mobility is discernible.

If this scenario is at work, an immediate question is raised about the cause for the mobility difference. The first thing to be considered is the effective mass. A larger effective mass would result in smaller carrier mobility. In YSb, although the electron effective mass is anisotropic along different directions, only a moderate difference is observed between electron and holes. A second possibility is a different scattering rate. A strong scattering in holes would reduce the hole mobility. This scenario is consistent with an inter-pocket (between α and β pockets) small q scattering, since these two pockets are close in the momentum space. Further study regarding the particular scattering mechanism would be very interesting.

Another issue to be considered is the MR in extremely large magnetic field. According to the two-band model, any deviation from perfect carrier compensation would eventually lead to a saturation of the MR when the magnetic field is sufficiently large. However, two other factors may also come into play. First, the above process is eased by the substantial difference in carrier mobility, such that the expected saturation from two-band model would only happen with a very large magnetic field. Second, when the magnetic field becomes very large, the two-band model would break down and the Landau level physics might take place and enhance the MR. Moreover, experimental evidence has been reported that the Fermi surface might be modified by the external magnetic field and become favorable to XMR if the original electronic structure is close to a carrier compensation status[3,29]. To what degree the above factors are intertwined with each other and contribute to the XMR in YSb at extremely large magnetic field remains to be explored.

In summary, by investigating the electronic structure of YSb, we provide spectroscopic evidence for the existence of XMR in a non-magnetic semimetal which lacks topological protection and perfect electron-hole compensation. A cooperative action of the moderate carrier compensation and the substantial mobility difference in YSb might provide a new pathway toward realizing XMR in non-magnetic semimetals.

We thank E. Y. Ma, S. N. Rebec, and X. Dai for useful discussions. The work at SLAC and Stanford is supported by the US DOE, Office of Basic Energy Science, Division of Materials Science and Engineering. ALS and SSRL are operated by the Office of Basic Energy Sciences, US DOE, under contract Nos. DE-AC02-05CH11231 and DE-AC02-76SF00515, respectively. Work at Argonne (sample growth and characterization studies) is supported by the US DOE, Office of Basic Energy Science, Materials Science and Engineering Division. J. J. and C. C. H. acknowledge support from the NRF, Korea through the SRC center for Topological Matter (No. 2011-0030787).

[#]*These authors contributed equally to this work.*

*To whom correspondence should be addressed: zxshen@stanford.edu*

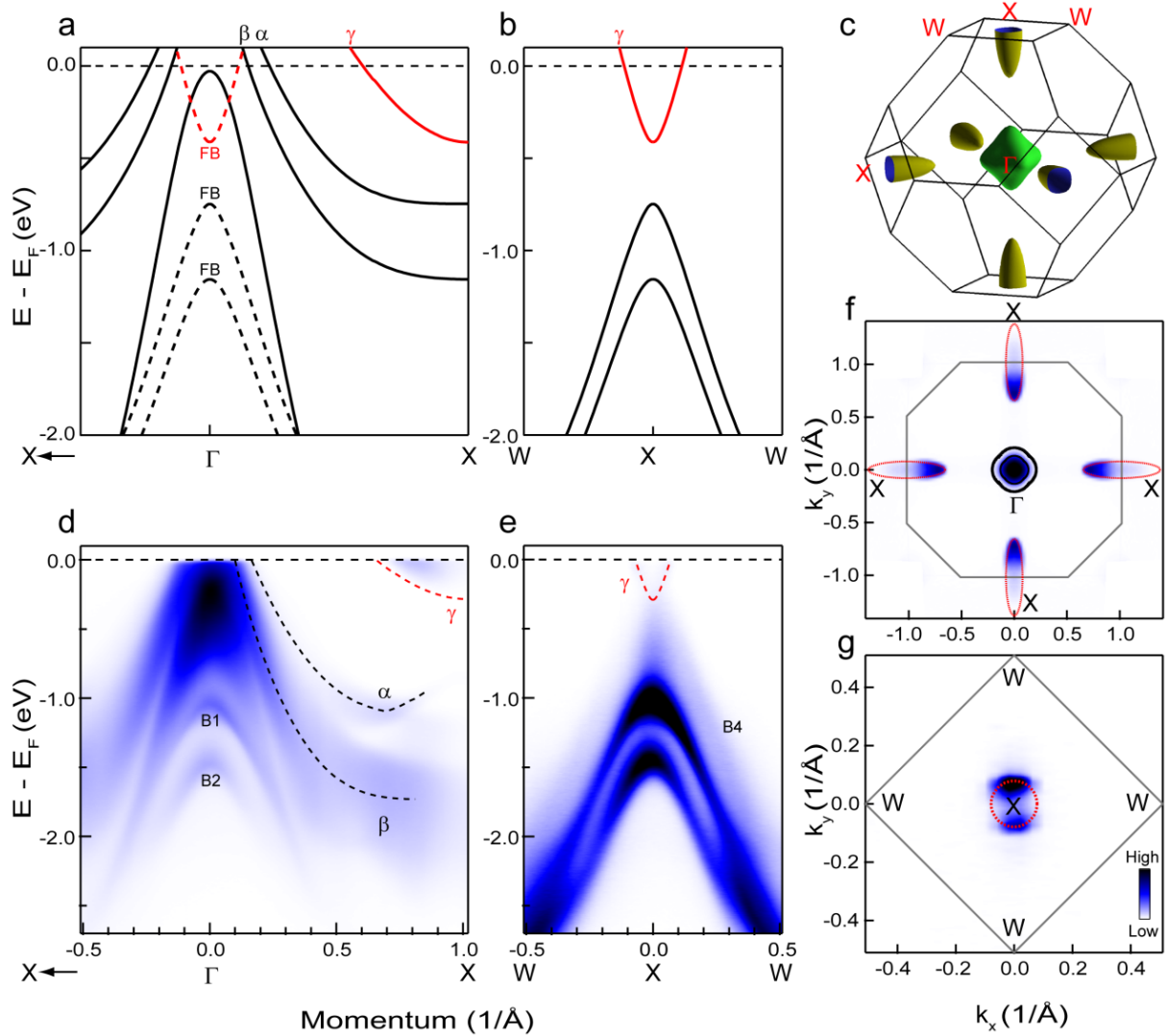

FIG. 1: Fermi surface and band structure of YSb. (a),(b) Calculated band structure along X-Γ-X and W-X-W. The conduction and valence bands are plotted in red and black respectively. There are two hole-like bands (α, β) and one electron-like band (γ) crossing the Fermi level. Dashed lines indicate selected folded bands (FB) from the calculation. The calculated chemical potential has been shifted up by 50 meV to fit the experiment. (c) Calculated Fermi surface shown in the 3D FCC Brillouin zone. (d),(e) Photoemission intensity plot of the band structure along X-Γ-X and W-X-W probed with the photon energy of 53 eV and 90 eV respectively. Various bands are guided by the dashed lines. A weak hole-like band feature (marked as B4) is also seen in (e) whose band top locates at ~300 meV below Fermi level. (f) Symmetrized Fermi surface mapping in the $K_x$-$K_y$ plane at the zone center ($K_z$=0, probed with 53 eV photons). (g) Fermi surface mapping in the $K_x$-$K_y$ plane at the zone boundary [top surface of the 3D Brillouin zone in (c), probed with 90 eV photons].

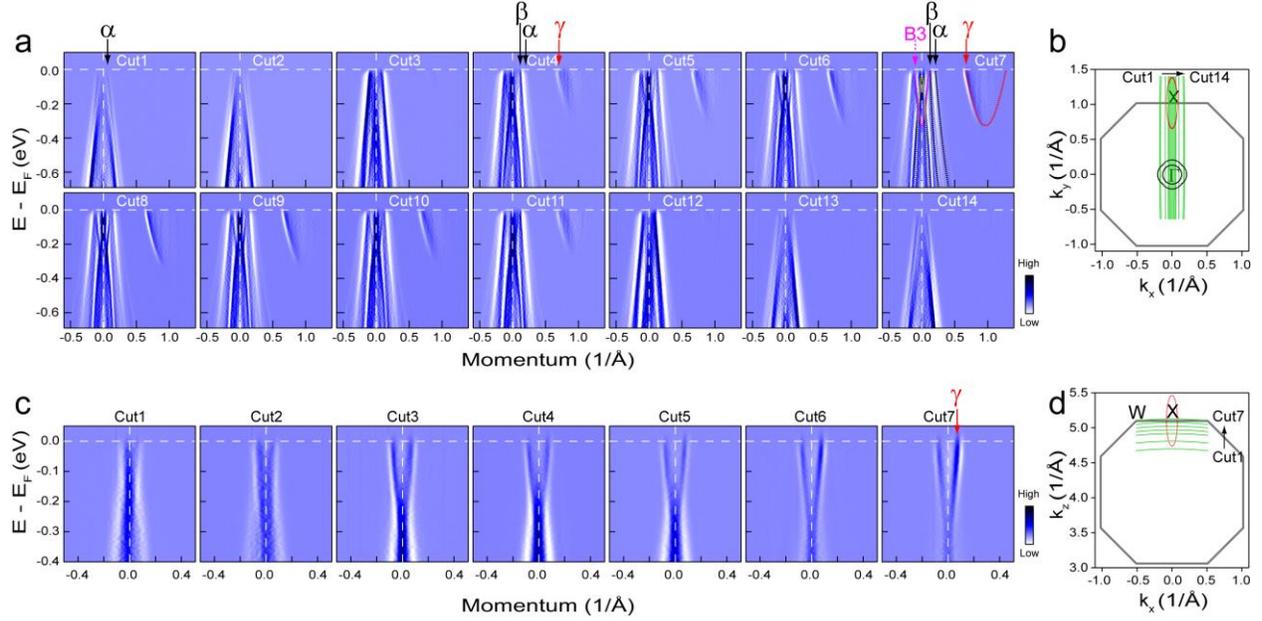

FIG. 2: Detailed band evolution near the Fermi pockets. (a) Momentum second derivative band structure near the electron and hole pockets in the $K_x$-$K_y$ plane at $K_z=0$, measured from cut 1 (top-left panel) to cut 14 (bottom-right panel). The black and red dashed lines in the top-right panel show schematically the hole-like bands (α, β) and electron-like band (γ) respectively. The pink and yellow dashed lines mark the additional bands near Γ. The location of the momentum cuts are shown in (b). (c) Band evolution (momentum second derivative images) near the electron-like pocket around X, measured in the $K_x$-$K_z$ plane at $K_y=0$ from momentum cut 1 to cut 7 in (d).

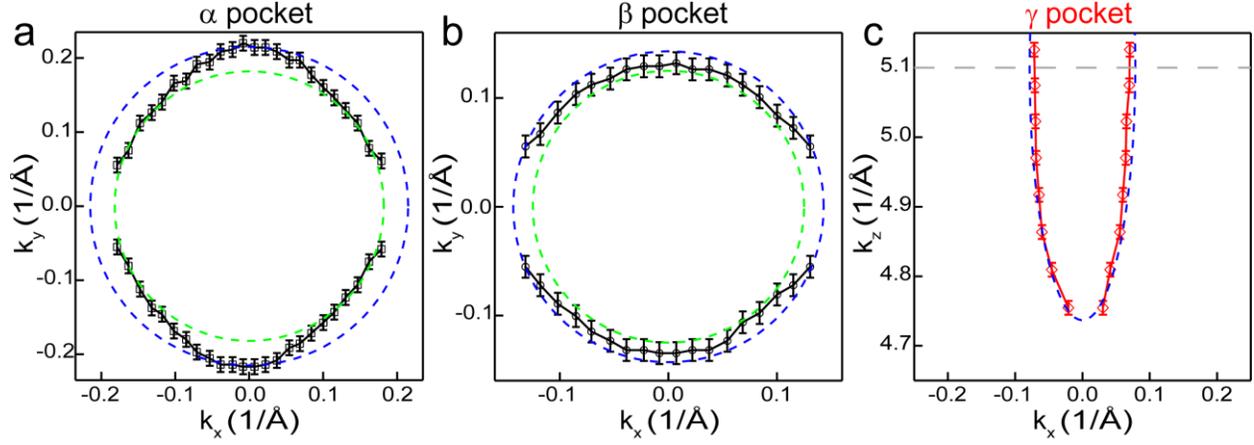

FIG. 3: Electron and hole Fermi pockets in YSb. (a),(b) Fermi surface of the α and β hole pockets in the $K_x$-$K_y$ plane at $K_z$=0, extracted from Fig. 2a. Two spheres with different radii are used to estimate the volume of each hole pocket, whose projection on the same $K_x$-$K_y$ plane is shown by the green and blue dashed lines, respectively. (c) Fermi surface of the γ electron pocket in the $K_x$-$K_z$ plane at $K_y$=0, extracted from Fig. 2c. The blue dashed line shows schematically the projection of an ellipsoid on the same $K_x$-$K_z$ plane, which is used to estimate the volume of γ pocket. Error bars reflect the uncertainty in determining the Fermi momenta.

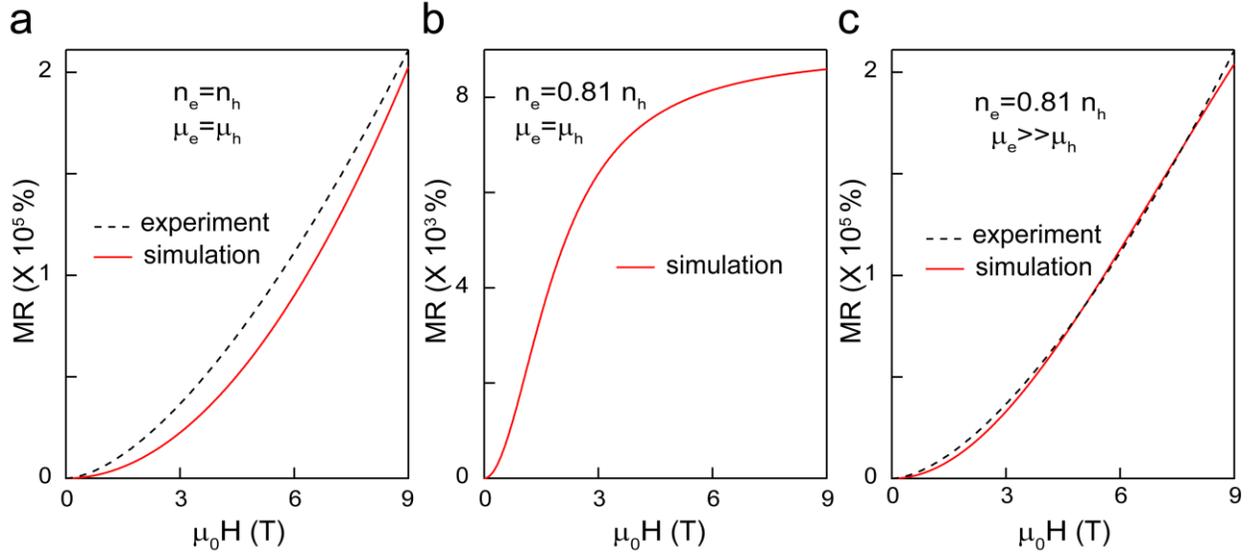

FIG. 4: The cooperative action of a moderate carrier compensation and a substantial difference between electron and hole mobility. (a) Simulated MR driven by perfect carrier compensation ($n_e=n_h$, red curve). The same carrier mobility [$\mu_e=\mu_h=5 \times 10^4$ cm$^2$(Vs)$^{-1}$] is used for the simulation. The black dashed curve represents the real MR measured by experiments. (b) Simulated MR when the carrier concentration deviates from perfect compensation [$n_e/n_h$=0.81, $\mu_e$ and $\mu_h$ are the same as those in (a)]. (c) Simulated MR using $n_e/n_h$=0.81, but with $\mu_e=9 \times 10^5$ cm$^2$(Vs)$^{-1}$ and $\mu_h=3.5 \times 10^3$ cm$^2$(Vs)$^{-1}$.

*Supplementary Information for*

**Distinct Electronic Structure for the Extreme Magnetoresistance in YSb**

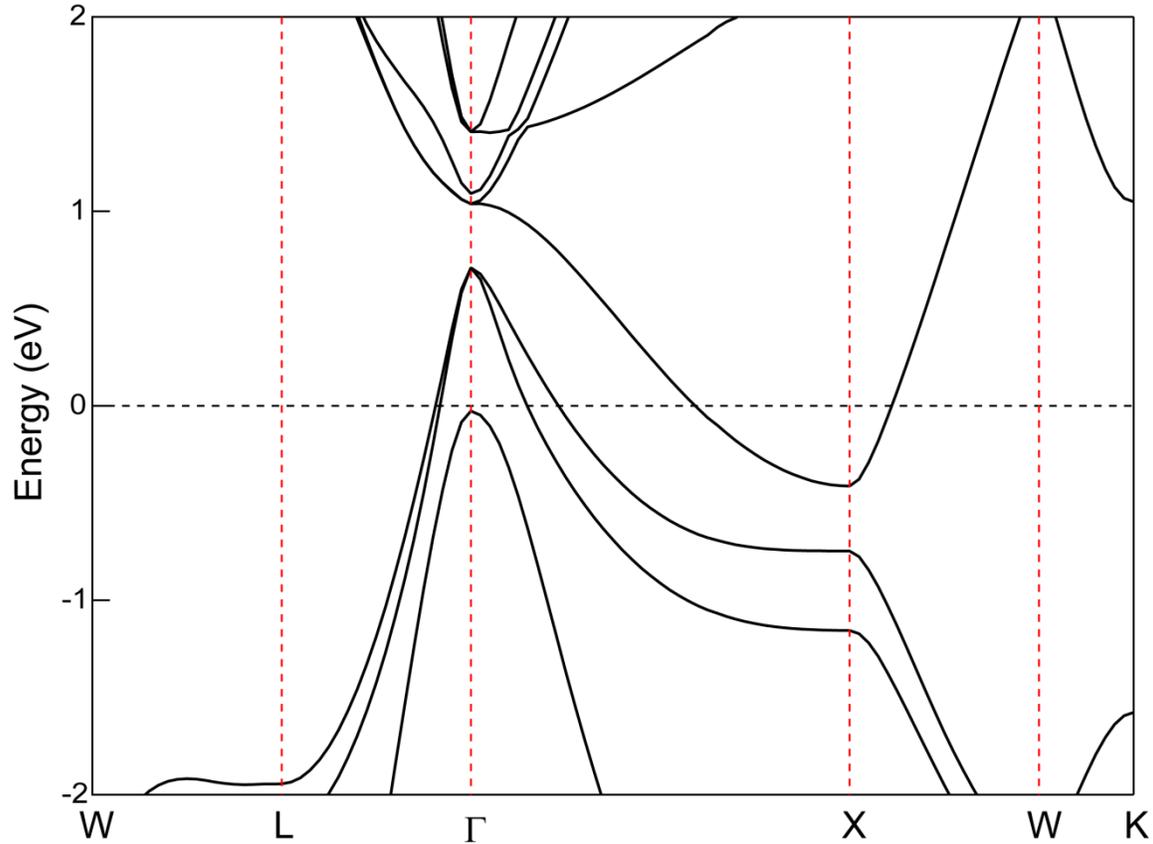

Supplementary FIG. 1: *ab initio* electronic structure calculations for YSb with a FCC crystal structure. Calculated band dispersion along high symmetry momentum directions. The calculations have been carried out with the WIEN2k code package using the standard PBE-GGA exchange-correlation potential. Spin-orbit coupling and mBJ potential are incorporated for improved estimates of the dispersion and band gaps. The calculated chemical potential has been shifted up by 50meV to fit the experiment. No band inversion is observed near Fermi level.

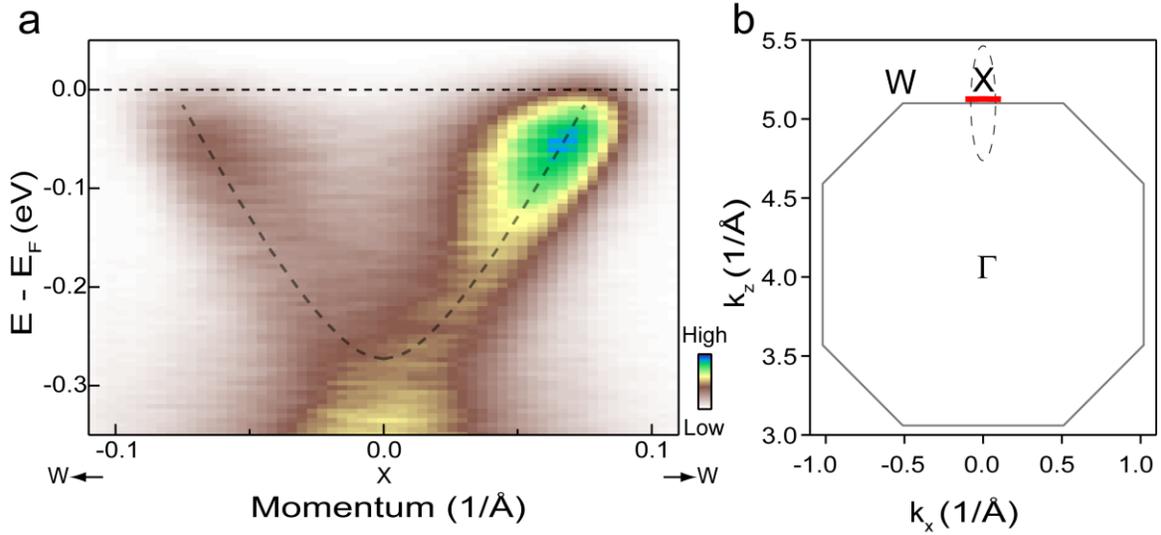

Supplementary FIG. 2: Parabolic dispersion of the electron-like band. (a) Raw photoemission intensity plot for the electron-like band dispersion around X. Dashed line is a guide for the eye, indicating the parabolic dispersion of the band. (b) Momentum location of the cut, same as Cut7 in main Fig. 2d but with a smaller momentum window.

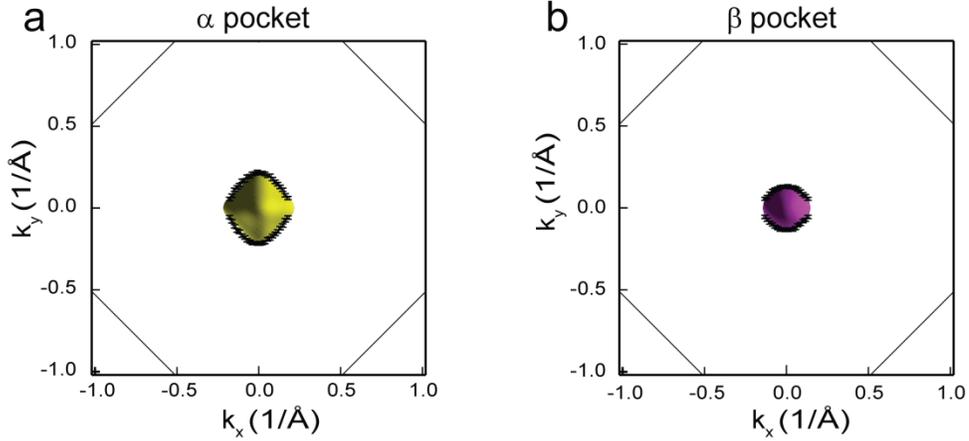

Supplementary FIG. 3: Estimation of the hole pockets volume by comparing with the DFT calculations. (a) 2D projection of the calculated 3D α hole pocket (yellow) and the measured 2D Fermi surface of the α pocket in the $K_x$-$K_y$ plane at $K_z=0$ (black). The Fermi level of the calculation has been shifted, such that the 2D projection best fits the experiment. Then, the calculated 3D hole pocket gives a good estimation of the real volume of the α pocket. (b) The same as (a) but for β hole pocket (2D projection shown in purple).

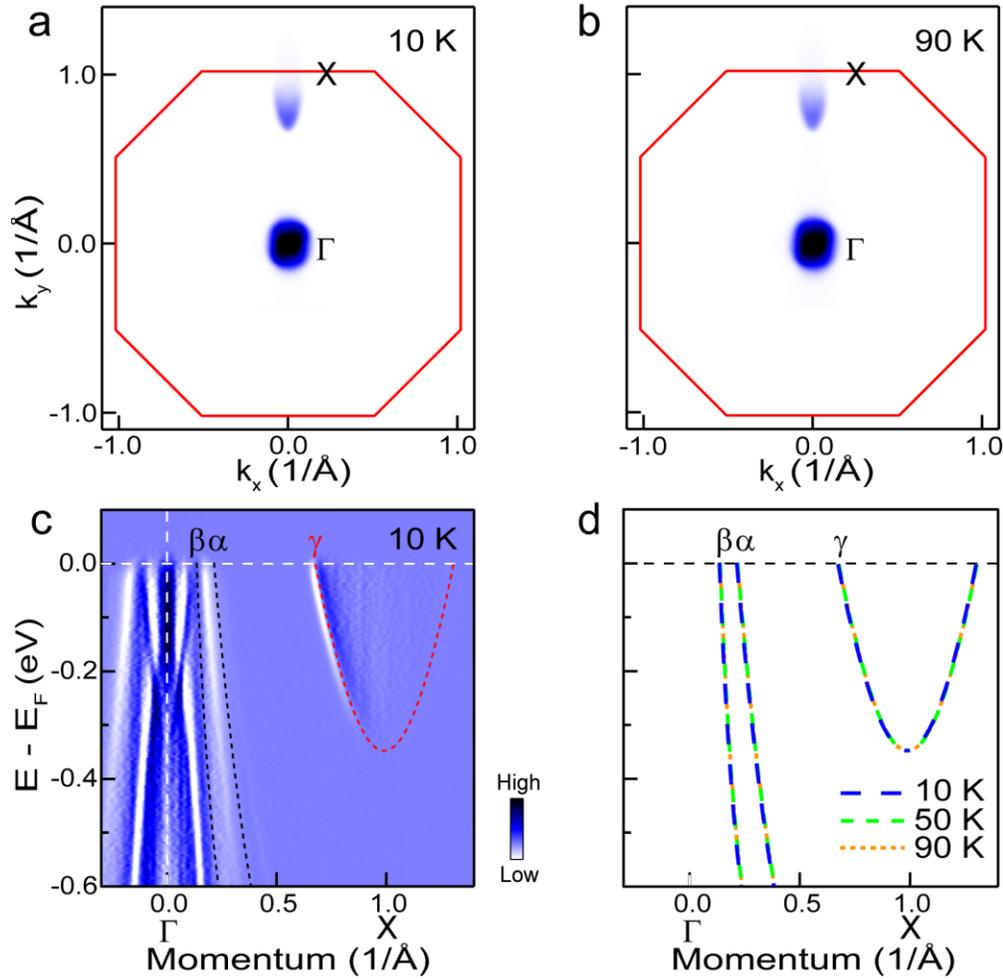

Supplementary FIG. 4: Temperature dependence of the electron and hole pockets. (a),(b) Fermi surface mapping in the $K_x$-$K_y$ plane ($K_z=0$) at 10 K and 90 K. The size of the electron and hole pockets shows little change with temperature. (c) Momentum second derivative images of the band structure along Γ-X at 10 K. (d) Band dispersions extracted from the momentum second derivative images at 10 K, 50 K and 90 K.

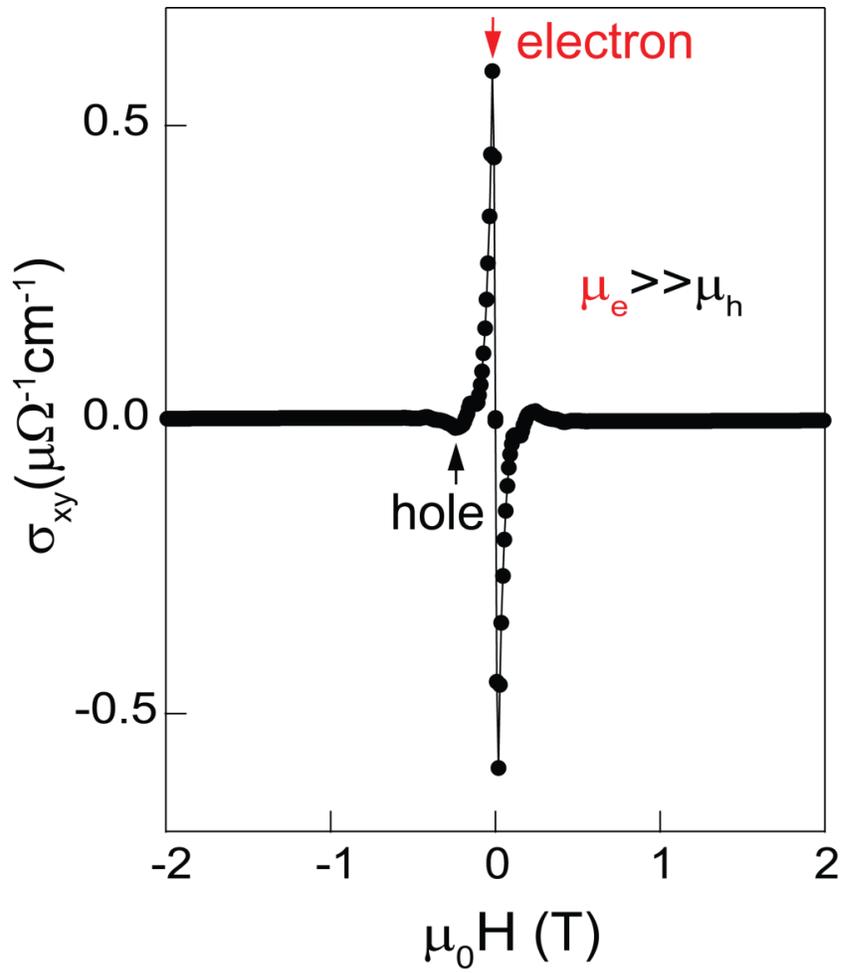

Supplementary FIG. 5: Hall conductivity $\sigma_{xy}$ versus magnetic field. The peak (marked by red arrow) and dip (marked by black arrow) locate the electron and hole mobility respectively.